# Developer Experience: Concept and Definition

Fabian Fagerholm, Jürgen Münch
*Department of Computer Science, University of Helsinki*
*P.O. Box 68, FI-00014 University of Helsinki, Finland*
*fabian.fagerholm@helsinki.fi*

*Abstract*—New ways of working such as globally distributed development or the integration of self-motivated external developers into software ecosystems will require a better and more comprehensive understanding of developers' feelings, perceptions, motivations and identification with their tasks in their respective project environments. User experience is a concept that captures how persons feel about products, systems and services. It evolved from disciplines such as interaction design and usability to a much richer scope that includes feelings, motivations, and satisfaction. Similarly, developer experience could be defined as a means for capturing how developers think and feel about their activities within their working environments, with the assumption that an improvement of the developer experience has positive impacts on characteristics such as sustained team and project performance. This article motivates the importance of developer experience, sketches related approaches from other domains, proposes a definition of developer experience that is derived from similar concepts in other domains, describes an ongoing empirical study to better understand developer experience, and finally gives an outlook on planned future research activities.

*Keywords*-Developer experience; software development environment; high-performing teams; human factors; software psychology

## I. Introduction

Software development is an inherently human-based, intellectual activity [1]. Many studies show that human factors are the most important factors for software development in many different development environments, both in terms of performance [2], [3] and quality [4], [5], [6], [7]. The studies clearly indicate the strong reliance of software project success on humans, while tools and methods only amplify the productivity of highly skilled and well-coordinated development teams. DeMarco notes that "companies that sensibly manage their investment in people will prosper in the long run" [8].

The increase of global software development (GSD) implies that an increasing number of software developers and software development teams are doing distributed software development [9], [10], [11]. In consequence, there is a strong indication that human factors are getting even more important and need to be better understood. New aspects such as communication across cultural, geographical, and temporal distances require significant attention. Software development environments are complex social systems, and flaws in communication and coordination will lead to failure of the development effort [12]. Research on GSD has shown that lack of trust, difficulties in communication, and lack of identification with project goals negatively impact the success of projects [11], [13], [14]. These difficulties exist already in a single multi-site company distributed geographically, and become more pronounced when development is also distributed over a network of collaborating organizations. The importance of complex factors such as trust, communication, and identification with goals is becoming more visible.

Taking a closer look at human factors in software development reveals, for instance, that a multitude of factors influence developer productivity (Nelson-Jones law). Summarizing research on productivity factors, Endres et al. come up with the conclusion that the number of factors ranges in the hundreds or thousands [1]. In addition, as DeMarco and Lister point out, there are hundreds of ways to influence productivity, and most of them are non-technical [3]. Overemphasis on productivity is the best way to lose it [1].

One idea that is proposed in this article is the definition of *developer experience* (DE$^x$). This concept is intended to a) abstract from the huge variety and quantity of human characteristics and factors without ignoring either, and b) to be sufficiently intuitive and concrete for practitioners. The main purposes the authors aim at with the concept of developer experience are to help practitioners to better understand, analyze, design and improve project environments with respect to developers' perceptions and feelings. The idea was influenced by the "user experience" concept (UX), a similar concept that aggregates relevant aspects, is intuitive, and helps organizations to analyze, design or improve products or services. This article describes developer experience as a concept that captures how developers think and feel about their activities within their working environments, with the assumption that an improvement of the developer experience has a positive impact on software development project outcomes. Section II sketches related approaches from other domains. Section III proposes a definition of developer experience, and Section IV describes current and planned future activities.

## II. RELATED APPROACHES

Several research approaches exist that address "experience" or similar concepts. In general, "experience" refers to both immediately perceived events as well as the memories of events and the knowledge gained by interpreting and reflecting on remembered events. Human experience is necessarily subjective: as our ability to process data is limited, we maintain an individual mental state of reality, which we use to interpret new data. Several types of experience exists: among the most relevant with respect to developer experience are the concepts of "user experience", "customer experience", and "brand experience". Other research approaches aim at better understanding, capturing, and modeling human factors in (software) projects, for instance, models for high-performing teams or simulation approaches. Finally, some approaches focus on creating "experience". In the following, these approaches are shortly sketched.

*User Experience (UX)* covers "a person's perceptions and responses that result from the use or anticipated use of a product, system, or service" [15]. This includes perceived product properties such as value, desirability, and usefulness. UX distinguishes between the verb *experiencing* – an individual's stream of perceptions, interpretations of those perceptions, and resulting emotions during an encounter with a system – and the noun *an experience* – an encounter with a system that has a beginning and an end [16], [17]. UX also recognizes *co-experience*, *shared experience*, or *group experience*, where the experience is socially constructed by several people. The definition of UX is still evolving and various aspects are being differentiated. However, the UX concept is sufficiently clear to draw parallels to the experience of software developers.

*Customer experience* occurs over time when a customer interacts with a supplier of goods or services [18]. It can also be used to mean an individual experience over one transaction: the customer experience concept includes both the cumulative experience and episodic experiences. It includes the experience of both a product or service, and the process during which the customer interacts with the supplier.

*Brand experience.* In marketing, a brand is a "name, term, design, symbol, or any other feature that identifies one seller's good or service as distinct from those of other sellers" [19]. Brand experience is conceptualized as subjective, internal consumer responses and behavioral responses evoked by brand-related stimuli that are part of the brand's design and identity, packaging, communications, and environments [20]. In creating brand experience, the goal is to develop or align the expectations behind the brand experience to create an impression that the brand has qualities and characteristics that make it unique or special. A brand is therefore one of the most valuable elements in an advertising theme.

*Models for high-performing teams.* The relationship between team performance and project success is unclear and can be understood in different ways [21], [22]. Individual performance is moderated by several factors, and team performance is a function of individual performance, group dynamics, and context factors. Psychophysical and -social needs influencing individual performance can be roughly divided into motivator and hygiene factors [23]. Motivator factors are related to the work itself and can increase performance, while hygiene factors are unrelated to the work itself and can result in dissatisfaction if they are missing (e.g. salary).

A broad division of factors influencing individual performance is individual characteristics (e.g. need for variety), internal controls (e.g. personality), and external moderators (e.g. career stage) [24]. More specifically, several groups of factors have been observed to affect performance: i) task characteristics, such as problem solving [24], technical challenge [24], [25], [26], and the nature of the job itself [27], [28]; ii) characteristics related to self-development and outside visibility of work, such as opportunities for advancement and growth, working to benefit others [24], recognition, opportunities for achievement [25], [26], increased responsibility [29], and senior management support [30], [31], [25], [26]; and iii) material and safety factors, such as salary, benefits, and job security [25], [26].

Theory on high-performing teams suggests that high morale, clear purposes through an open atmosphere, communication and honest feedback within the team are key factors [32]. Other important team-level performance factors are high technical competence, fully documenting work, and sharing knowledge with the team [26], interdependencies between team members, team synergy, and the ability to share a common vision for the software to be developed [33]. High-performing teams are reportedly proud of their high technical competence and confidence, and the flexible and adaptable attitudes of the members [26]. They value the ability to communicate, listen, give relevant feedback, see the "big picture", and be a good team worker, but dislike hiding problems and work. Trust is key to the cohesive team [14]. Diverse knowledge positively affects team performance while diversity in personal values may affect it negatively [34].

*Simulation modeling* abstracts some particular process in software development, maintenance, or evolution [35]. Software process simulation models mainly capture the dynamics of software development and can be used, for instance, to aid decision-making, risk reduction, and management at strategic, tactical, and operational levels. Simulation models may include human factors-related variables, such as hiring rate, staff turnover rate, capabilities, motivation, and training provided. Many simulation models for software development processes have been proposed. Some simulation models for capturing human behavior exist [36]. However, most often existing simulation models are not detailed enough to capture human characteristics.

Several other approaches emphasize the creation of experience: *Experience design*, for instance, is the practice of designing products, processes, services, events, and environments with a focus placed on the quality of the user experience and culturally relevant solutions, with less emphasis placed on increasing and improving functionality of the design [37]. Another example, Dewey's theory of *art as an experience*, says that the entire artistic process should be considered important, not only the physical art-object [38]. Here, the object itself is not the fundamental goal, but rather the development of an "experience" which recaptures some aspect of life. Similarly, the creation of software can be considered an experience, and methods that govern it shape that experience.

## III. DEVELOPER EXPERIENCE

This section presents a definition of *developer experience* ($DE^x$). The definition is influenced by the UX concept. We assume that several factors influence $DE^x$, which in turn affects outcomes of software development projects. The word "developer" refers here to anyone who is engaged in the activity of developing software, and "experience" refers here to involvement, not to being experienced, although the two are interlinked.

UX has evolved beyond user interface design. Practitioners and researchers know that it is not sufficient to focus only on the user interface and on avoiding usage defects, increasing robustness, and ensuring safety. The user perspective has been shifted first to efficiency and ease of use, then to the question of appropriate use and fitness for purpose, and finally to considering the entire experience of using a (software) product or service. This progression can be thought of as a maturation process in the human-computer interaction field. In Table I, we draw a parallel to the developer perspective, where the end goal is not to use but to create a product or service. Here, the basic level is exemplified by prescriptive process models, the efficiency level by descriptive and adaptive process models, the appropriateness level by a detailed understanding of the process-product relationship in a specific context, and the final level by the entire experience of being a software developer and carrying out software development activities.

As noted, software development is an intellectual activity, which rests on the capabilities of the mind, requiring both thought and motivation to carry out. In psychology, the concept of mind is commonly divided into cognition (attention, memory, producing and understanding language, problem-solving, decision-making), affect (feeling, emotion), and conation (impulse, desire, volition, striving). We explicitly include in $DE^x$ not only affective aspects, but also cognitive, conative, and social aspects of experience. Since the end goal in the developer perspective is to create software, it is especially important to consider how thought and feeling is turned into intentional action, and how group work should be systematically organized to support this.

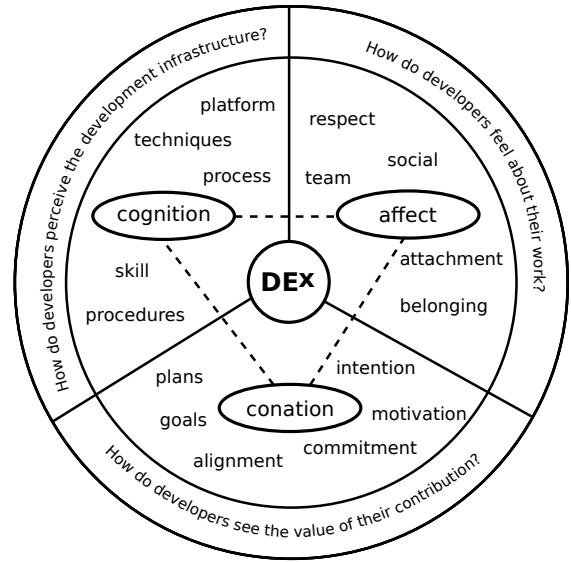

Figure 1. Developer Experience: Conceptual framework.

$DE^x$ consists of experiences relating to all kinds of artifacts and activities that a developer may encounter as part of their involvement in software development. These could roughly be divided into experiences regarding i) development infrastructure (e.g. development and management tools, programming languages, libraries, platforms, frameworks, processes, and methods, ii) feelings about work (e.g. respect, attachment, belonging), and iii) the value of one's own contribution (e.g. alignment of one's own goals with those of the project, plans, intentions, and commitment). Figure 1 shows the concept of $DE^x$ as an interaction between cognitive, affective, and conative factors. Each dimension of $DE^x$ consists of a multitude of complex sub-factors. The cognitive dimension consists of factors that affect how the developers perceive their development infrastructure on an intellectual level. This includes concrete interactions with development tools and execution of a software process. Perceiving these in a positive light is likely to contribute to better $DE^x$. The affective dimension consists of factors that influence how developers feel about their work. Respect and belonging are social factors that work to create a feeling of security. Attachment to persons, teams, or even habits of work also belong to this dimension. Positive feelings in general can be an important factor in good $DE^x$. The conative dimension consists of factors that affect how developers see the value of their contribution. Intentional, planned activity with personal goals that are properly aligned with the goals of others is likely to increase the sense of purpose, motivation, and commitment, and thus positively affect $DE^x$.

TABLE I
TRANSFER FROM USER EXPERIENCE TO DEVELOPER EXPERIENCE: FOCUS AREAS AND END GOAL

| Focus | User perspective | Developer perspective |
| --- | --- | --- |
| Positive experience + appropriate use + efficient use | User Experience | Developer Experience |
| Appropriate use, fitness for purpose + efficient use | User-centered design | Understanding of process-product relationship for a specific context |
| Efficiency and ease of use | Usability | Descriptive process models, adaptive process models |
| Avoid usage defects, increase robustness, safety | User interface design | Prescriptive process models |
| **End goal** | Use product/service | Create product/service |

$DE^x$ may be important in several areas of software development. For example, in software process improvement, it could give valuable input for analyzing and adjusting processes. In software project management, it could offer means to evaluate plans and goals with respect to their alignment with developers' motivation and commitment. For maintaining development team performance, it could offer insight into factors that affect sustainable team work. In designing development platforms and environments, e.g. when a platform provider attempts to grow an ecosystem of developers to create applications and services, $DE^x$ may offer means to design the development experience so that platform and ecosystem is more attractive to developers.

## IV. ONGOING AND FUTURE WORK

The presented model is mainly based on literature reviews and transferring ideas from related concepts and other domains as well as experiences of the authors. However, we aim at developing a sound empirically-based model for developer experience. In order to create such a model, a first study has been started that focuses on practitioner's understanding of the phenomenon "developer experience". The goal of this first study is to identify practitioners' conceptions of developer feelings and development experience when working in a specific project environment. The research question is: How do practitioners characterize a software developer's experience in a specific development environment and what kind of impact factors on this experience do they consider as relevant in this respective environment? The underlying motivation for the study is to find common perceptions among practitioners, or at least to analyze the degree of agreement on characteristics and impact factors of developer experience. A second motivation is to identify differences and variations in practitioners' conceptualization of developer experience and to find out how they could be explained (e.g. by different roles, different project settings, different cultures).

The study is conducted in two contexts: The first context is a set of close-to-industry student projects in the Software Factory laboratory at the Department of Computer Science, University of Helsinki. The study uses a number of data collection and analysis methods. Basic background questionnaires are used to gain demographic data about project participants. Video analysis is used to analyze specific project events in detail and to gain an understanding of how individual differences are visible in certain situations. Focus group interviews are used to do postmortem analyses of the projects and gain access to rich, qualitative explanations by team members. The second study is an in-depth interview study in five companies, in which we interviewed practitioners working in different areas of software development, software product and service design, and management. The research method of the interview study is based on thematic interviews to collect data, and an Affinity diagram as a sense-making tool to analyze the data. An initial content analysis of this data provided a rich, qualitative understanding of $DE^x$.

Future work focuses on a refinement or modification of the initial model presented in this article based on the findings of the first study. Afterwards, a concept evaluation including a more detailed analysis of the construct validity is planned. In the long run, we plan to select sub-areas of $DE^x$ for more detailed studies. We plan to develop measurement instruments for them and get a better understanding of the impact factors in specific contexts. Finally, we plan to perform empirical studies to examine the effect of key characteristics of developer experience on aspects such as project performance. In addition, complementary guidelines and analysis tools are planned to help practitioners to design developer experience in a way that it supports higher-level organizational goals and business strategies.